# Self-similar solutions in cylindrical magneto-hydrodynamic blast waves with energy injection at the centre

A. Gintrand ★ and Q. Moreno-Gelos
*Extreme Light Infrastructure ERIC, ELI Beamlines Facility, Za Radnici 835, 252 41 Dolní Břežany, Czech Republic*

**ABSTRACT**
The evolution of shocks induced by massive stars does not depend only on the ambient magnetic field strength, but also on its orientation. In the present work, the dynamics of a magnetized blast wave is investigated under the influence of both azimuthal and axial ambient magnetic fields. The blast wave is driven by a central source and forms a shell that results from the accumulation of interstellar matter behind the shock front. A similarity form of the ambient magnetic field and a cylindrical geometry of the blast wave are assumed to obtain self-similar solutions. The model is studied separately for both azimuthal and axial magnetic field and applied to stellar wind bubbles and supernova remnants respectively, using 1D numerical simulations. We found that the magnetized blast wave differs from the self-similar case without an ambient magnetic field. The forward shock front goes slower in the azimuthal case and faster in the axial one. For both tangential orientations, the thickness of the shell increases with the magnetic strength. In the azimuthal case, the thermal energy can be converted to magnetic energy near the inner boundary of the shell. Thus, the temperature drops and the magnetic field increases at the tangential discontinuity of the stellar wind bubble. In the axial case of a supernova remnant, the numerical solution always follows a special curve in the parameter space given by the self-similar model.

## 1 INTRODUCTION

Magneto-Hydrodynamic (MHD) shocks are ubiquitous in the universe. The study of those shocks is important to understand the structure and evolution of many astrophysical objects such as Supernova Remnants (SNRs), Stellar Wind Bubbles (SWBs), and jets. The SNR results from a huge amount of energy released in the interstellar medium (ISM). During the propagation of the Blast Wave (BW), the ISM material is accumulated behind the shock to form a dense shell. When the mass of the shell becomes comparable to the mass of the ejecta from the supernova, the shock starts to decelerate. This can be described by the so-called self-similar solution of Sedov–Taylor (Sedov 1959; Taylor 1950). Thus, the radius of the shock $R_{sh}$ follows the evolution $R_{sh} \propto t^\alpha$, where $\alpha = 2/5$ (in spherical geometry). For the SWBs, the shock is driven by a constant flow of stellar material coming from the central star that acts as a piston and drives the shock through the ISM. Similarly, when the ISM matter swept up by the forward shock is comparable to the mass of the interior stellar wind, the radius of the SWB starts to decelerate following the self-similar expansion $R_{sh} \propto t^\alpha$, where $\alpha = 3/5$ (Ryu & Vishniac 1988; Koo & McKee 1992). This self-similar expansion is also found in the propagation of supersonic jets (Falle 1991) where the constant input of kinetic energy to the bow shock is given by the jet material. However, the morphology of the BW can be altered by the presence of an ambient magnetic field. In MHD shocks, the tangential component of the magnetic field is responsible for the departure from the pure hydrodynamic case. This tangential magnetic field $B$ can be divided into two categories according to their azimuthal $B = B_\phi$ or axial $B = B_z$ direction in the cylindrical coordinates. It will be demonstrated that those two magnetic fields, although they are both perpendicular to the forward shock propagation, will have different effects on the BW dynamics. First, in the ISM, the azimuthal magnetic field $B = B_\phi$ is usually generated by a central source. Indeed, during the early stage of the SWB propagation, the BW is affected by an azimuthal magnetic field generated by the rotating central star (Weber & Davis Jr 1967; Usov & Melrose 1992). The magnetic field is of the order 10 G at the surface and decreases as $B_\phi \propto 1/r$ where $r$ is the radial distance from the star. For older SWBs, $B_\phi$ can become lower than the ambient interstellar magnetic field of the order $10\,\mu G$. This brings us to the second type of magnetic field which is the axial field $B = B_z$ that can be locally encountered in the ISM during the propagation of the BW. In this case, the morphology of the nebula changes to a barrel-like shape where the matter is more confined in the direction perpendicular to the axial magnetic lines due to the magnetic pressure (Heiligman 1980; Soker & Dgani 1997; van Marle, Meliani & Marcowith 2015). This barrel shape behaviour also occurs during the evolution of SNRs. Moreover, during the radiative stage of old SNRs, the presence of an ambient magnetic field can alter the dynamics of the thin radiative shell. Indeed, the tension of magnetic lines can increase the thickness of the shell, decrease the compression of the density and prevent the further deceleration of the shock front radius to the so-called Pressure Driven Snowplow stage $R_{sh} \propto t^\alpha$, where $\alpha = 2/7$ (Blondin et al. 1998; Bandiera & Petruk 2004; Petruk et al. 2018). In astrophysical jets, the propagation is also collimated by the presence of a magnetic field generated by a central source like active galactic nucleus (AGN) and protostars. The quasi-cylindrical collimation of the jet can be explained by the presence of both the azimuthal and axial magnetic fields. Close to the source, the azimuthal magnetic field is generated by the rotation of the central object. Also, the accretion disc surrounding the central source can generate a poloidal field where the lines at radial distance $r$ close to the centre are almost directed along the $z$-axis of the jet. This second magnetic field could be a better candidate for the collimation of the jet as it is not subject to the kink instability (Spruit, Foglizzo & Stehle 1997; Moll 2009). The effect of the ambient magnetic field has also been studied in laboratory experiments using laser and Z-pinch induced shocks. In the laser experiment of Mabey et al. (2020), the BW exhibits a spheroidal shape where the major axis is aligned with the ambient axial magnetic field. Also, in different geometry, the study of a cylindrical BW propagating in an axial magnetic field has been performed using a Z-pinch machine (Vlases 1963; Vlases & Jones 1968). It has been found that if the BW is expanding in a non-conducting gas, the shock will slow down in presence of an ambient magnetic field. However, if the BW is propagating in a pre-ionized medium, the magnetic field will speed up the forward shock.

---

★ E-mail: antoine.gintrand@gmail.com

In the following, we will model the effect of the two tangential magnetic fields on a cylindrical BW to conserve the radial symmetry of the configuration in both azimuthal and axial cases. Then, we will apply the model to astrophysical scales for both azimuthal and axial fields. By choosing a special form of the magnetic field, the self-similarity of the dynamics will be conserved. This has already been done for the special case of the Sedov–Taylor solution in cylindrical geometry by Greifinger & Cole (1962). The authors assume that this BW is embedded in an azimuthal magnetic field $B = B_\phi \propto 1/r$ generated by a central constant current. In the model of Vishwakarma & Yadav (2003), the same cylindrical BW propagates through an ambient axial magnetic field and an ambient density profile in power law of the radial distance. A similar analysis has also been performed for the propagation of a spherical BW using power-law spatial variations of the ambient density and azimuthal magnetic field (Rosenau & Frankenthal 1976; Rosenau 1977). In all the different models, the magnetic field compression slows the propagation of the fluid behind the shock. Thus, the thickness of the shell increases with the magnetic strength. However, the presence of an ambient density power law together with the magnetic field makes the comparison between the different MHD cases difficult as the inhomogeneous density profile also plays a significant role in the behaviour of the solution. In this paper, we will consider a constant ambient density medium by allowing the total energy to vary with time according to a power law (Dokuchaev 2002; Sanz, Bouquet & Murakami 2011). In this case, the special role of the magnetic field with a homogeneous ambient density medium will be determined from point explosion solutions like the one found in SNRs, to wind-driven bubbles and jets. The paper is organized as follows. First, the self-similar model is presented in Section 2. In Section 3, the self-similar solutions for the case of an azimuthal ambient magnetic field are determined, which reveals a cooling by transfer from thermal to magnetic energy close to the inner boundary of the shell. The second case of an axial magnetic field is performed in Section 4, where the physical solutions are only allowed in a special region of dimensionless parameters. In Section 5, the self-similar solutions are applied to study the dynamics of SWBs and SNRs in the presence of a power-law magnetic field according to astrophysical conditions. Although the magnetic field is not self-similar anymore, a good agreement is found between the self-similar model and the internal structure of the numerical solution. Moreover, the numerical solution moves in the parameter space while conserving the properties of the self-similar solution. Finally, the discussions and conclusions are presented in Section 6.

## 2 SELF-SIMILAR MODEL

In this section, we will consider the propagation of a cylindrical BW in a homogeneous ionized ideal gas with density $\rho_0$, pressure $P_0$, constant polytropic index $\gamma$ and infinite electrical conductivity immersed in an ambient magnetic field $B_0$. The BW is the result of a large amount of energy released in a thin cylinder with radius $r_0$ along the radial axis $r$ and infinite along the axial axis $z$. Unlike the point explosion problem that leads to the solution of Sedov–Taylor (Sedov 1959), we will assume that the energy injection at the centre is not necessarily instantaneous and punctual and varies with time following the power-law form $E_0 \propto t^\lambda$, where $\lambda \geq 0$. Also for the magnetic field, we suppose that it follows the form $|B_0| \propto 1/r^\delta$ where the exponent $\delta \leq 1$. These power-law assumptions are necessary to conserve the similarity property of the flow. Notice that the self-similar solutions presented in this section could have been found in the same way using a homogeneous ambient magnetic field that varies with a power law in time. For the magnetic field, the two particular cases of an azimuthal and an axial magnetic field are going to be studied separately. Depending on its direction and the value of the exponent $\delta$, the ambient medium needs an additional force (like gravitation) to be maintained in a static state. This cannot be supplied by an ambient thermal pressure that is supposed to be negligible compared to the internal pressure of the BW. Indeed, this assumption allows us to stay within the strong shock limit, so the Mach number $M \equiv \mathcal{S}/c_0 \gg 1$, where $\mathcal{S}$ is the shock velocity and $c_0 = (\gamma P_0/\rho_0)^{1/2}$ is the sound speed of the ambient medium. The jump equations for an ideal MHD shock with $M \gg 1$ propagating in a static medium, where the magnetic field is perpendicular to the direction of propagation, give the boundary conditions of the solution at the shock front (denoted *sh*),

$$\frac{\rho_{sh}}{\rho_0} = Q_1, \quad \frac{v_{sh}}{\mathcal{S}} = Q_2, \quad \frac{kT_{sh}}{m\mathcal{S}^2} = Q_3, \quad \frac{B_{sh}}{\sqrt{\mu\rho_0}\mathcal{S}} = Q_4, \quad (1)$$

where $v$ is the radial velocity, $T$ is the temperature, $k$ is the Boltzmann constant, $m$ is the average mass per particle, and $\mu$ is the permeability of the vacuum. The quantities $Q_i(\sigma, \gamma)$ are calculated in Appendix A and depend on the polytropic coefficient of the gas and the magnetic strength parameter $\sigma \equiv M_A^{-1}$, where $M_A = \sqrt{\mu\rho_0}\mathcal{S}/B_0$ is the Alfven Mach number. From the definition of a fast-magnetosonic shock $(1/M_A^2 + 1/M^2 < 1)$ and the assumption $M \gg 1$, we conclude that the parameter $\sigma$ is in the range $0 \leq \sigma \leq 1$. The limit case $\sigma = 0$ corresponds to the pure hydrodynamic case and $\sigma = 1$ corresponds to the strong magnetic regime where the discontinuity is not compressible anymore ($Q_1 \to 1$). Also, notice that the quantity $Q_4 = \sigma Q_1$, so that the compression of density and magnetic field are always the same. As mentioned before, we are looking for power-law self-similar solutions that can be found using the following transformations,

$$\rho = \rho_0 n(\xi), \quad v = \alpha \frac{r}{t} u(\xi),$$
$$kT/m = \alpha^2 \frac{r^2}{t^2} z(\xi), \quad B = \sqrt{\mu\rho_0} \alpha \frac{r}{t} b(\xi), \quad (2)$$

where the self-similar coordinate is $\xi = r/R_{sh}$. The radius $R_{sh}$ is the position of the shock front that follows the power law $R_{sh} = At^\alpha$. The exponent $\alpha$ is the decelerating parameter and $A$ is a positive constant. At this point, notice that the exponent $\delta$ of the ambient magnetic field must follow the relation $\delta = (1 - \alpha)/\alpha$. This comes from the fact that $B_0 \propto \mathcal{S}$ at the shock front (see equation 1). The following self-similar analysis will be focused on the shell of the BW. This shell is defined by the fluid between the shock front at $R_{sh}$ and the inner boundary $R_p \equiv \xi_p R_{sh}$, where $\xi_p$ is the self-similar inner boundary. This inner discontinuity acts like a piston that separates the shell from a hot interior bubble that pushes it through the ambient medium. In some particular cases, $R_p = 0$ and the shell will represent the entire volume of shocked ambient material from the shock front to the centre of the BW (see Section 3). Using the transformations (2), one can find from the total energy of the shell,

$$E(t) = \int_{R_p(t)}^{R_{sh}(t)} \left( \frac{1}{2}\rho v^2 + \frac{P}{(\gamma-1)} + \frac{B^2}{2\mu} \right) 2\pi r \, dr$$
$$= 2\pi \rho_0 R_{sh}^2 \mathcal{S}^2 \Pi_{tot}(\gamma, \alpha, \sigma) \propto t^\lambda, \quad (3)$$

where

$$\Pi_{tot}(\gamma, \alpha, \sigma) = \int_{\xi_p}^1 \left( \frac{1}{2}n\xi^2 u^2 + \frac{\xi^2 nz}{(\gamma-1)} + \frac{\xi^2 b^2}{2} \right) \xi \, d\xi \quad (4)$$

is the dimensionless total energy of the shell and

$$\lambda = 4\alpha - 2. \quad (5)$$

The quantities $P$ and $B^2/(2\mu)$ are the thermal and magnetic pressure respectively. From equation (4), one can recover two particular cases. First, when the total energy is conserved ($\lambda = 0$), the deceleration parameter $\alpha = 1/2$, which is the same value as the Sedov–Taylor solution (Sedov 1959). Secondly, when the injection of energy is supplied by a constant flux of energy at the centre ($\lambda = 1$), the decelerating parameter $\alpha = 3/4$. This situation can be representative of adiabatic SWB (Ryu & Vishniac 1988; Koo & McKee 1992; Pittard, Hartquist & Dyson 2001) and jets (Falle 1991; Kaiser & Alexander 1997). Then, one can also introduce the following energies,

$$E_{mag}(t) = \int_{R_p(t)}^{R_{sh}(t)} \frac{B_0^2}{2\mu} 2\pi r \, dr$$
$$= 2\pi \rho_0 R_{sh}^2 \mathcal{S}^2 \Pi_{mag}(\gamma, \alpha, \sigma), \quad (6)$$

and

$$\Delta E_{flu}(t) = I \int_{R_p(t)}^{R_{sh}(t)} B \, dr - I \int_0^{R_{sh}(t)} B_0 \, dr$$
$$= 2\pi \rho_0 R_{sh}^2 \mathcal{S}^2 \Pi_{flu}(\gamma, \alpha, \sigma), \quad (7)$$

where $\Pi_{mag}$ and $\Pi_{flu}$ are defined in Appendix B. The quantity $I \equiv 2\pi R_{sh} B_0(R_{sh})$ is the current that must be applied along the $z$-axis to generate a magnetic field $B_0$ at the shock front. For $\alpha = 1/2$ this current is constant, and for $\alpha > 1/2$ this current increases over time. Equation (6) is the total ambient magnetic energy swept by the BW. Equation (7) is the difference between the magnetic energy flux contained in the shell and the one contained in the BW's whole volume before the explosion. During the propagation of the BW, the energy $E$ contained in the shell and the energy $E_B$ contained in the interior bubble verify the following equation,

$$E(t) + E_B(t) = E_0(t) + E_{mag}(t) + \Delta E_{flu}(t). \quad (8)$$

Notice that the energy of the interior bubble is not affected by the ambient magnetic field where the energy supply comes from the central source only. By identification, we can deduce from equation (8) that the bubble energy also verifies $E_B \propto t^\lambda$. Thus the central energy source is divided between the interior bubble and the shell, and $E_B \equiv (1-\zeta)E_0$, where $\zeta$ is a constant in the range $0 \leq \zeta \leq 1$. Finally, the quantity $\zeta E_0$ supplied to the shell by the central source is independent of the ambient magnetic field, and from (8), its associated dimensionless energy reads,

$$\Pi = \Pi_{tot} - \Pi_{mag} - \Pi_{flu}, \quad (9)$$

where the BW radius verifies the following equation,

$$R_{sh}(t) = \left( \frac{\zeta}{2\pi\alpha^2} \frac{E_0(t)}{\rho_0} \frac{t^2}{\Pi(\gamma,\alpha,\sigma)} \right)^{1/4} \quad (10)$$

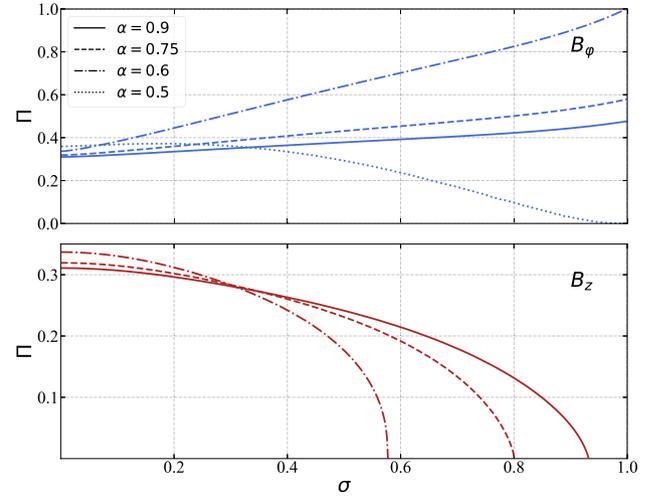

**Figure 1.** Variations of integral $\Pi$ versus the magnetic strength parameter $\sigma$ for $\alpha = 0.9$ (solid lines), $\alpha = 0.75$ (dashed lines), and $\alpha = 0.6$ (dotted-dashed lines) with $\gamma = 5/3$. The blue curves are associated with the azimuthal field $B_\phi$ and the red with the axial field $B_z$.

The computation of $\Pi$ is performed after integration of the equations in Sections 3 and 4. Equation (10) provides the effect of the magnetic field on the BW evolution according to the variations of $\Pi(\sigma)$. Only the parameter $\zeta$ remains unknown and needs to be computed numerically by solving the equations from the initial condition to the intermediate-asymptotic regime (Barenblatt 1996). In Fig. 1, one can see that except in the case $\alpha = 1/2$ (see below), the increase of an ambient azimuthal magnetic field $B_\phi$ results in the growth of the function $\Pi(\gamma, \alpha, \sigma)$ and the shock front deceleration. On the contrary, the shock velocity increases when increasing the axial ambient field. This behaviour will be recovered during the simulations of SWB and SNR where the solution is not self-similar anymore. In the axial case (bottom panel), each curve goes to zero for a specific value of $\sigma$. It will be demonstrated in Section 4 that no physical solution exists beyond this threshold value. For $\alpha > 1/2$, one can verify that $\Pi_{flu} = 0$, that is, the quantity $\Delta E_{flu} = 0$. Indeed, the total magnetic flux is conserved which means that all the energy flux swept by the BW is found in the shell. However, the special case $\alpha = 1/2$ is singular. As we will see in the next section 3, the solution is regular from the shock front to the centre of the BW and $R_p = 0$. Here, the quantity $\Pi_{flu} > 0$ represents the energy supplied after the explosion to maintain the constant current $I$ during the BW propagation (Greifinger & Cole 1962). In this case, the function $\Pi(\sigma)$ increases to a maximum around $\sigma \approx 0.175$ and then goes to zero when $\sigma \rightarrow 1$.

## 3 AZIMUTHAL MAGNETIC FIELD

In the case of an azimuthal magnetic field, the 1D cylindrical ideal MHD equations of mass, momentum, energy, and induction conservation read,

$$\partial_t \rho + \frac{1}{r}\partial_r(r\rho v) = 0, \tag{11a}$$

$$\rho(\partial_t v + v\partial_r v) + \partial_r\left(P + \frac{B^2}{2\mu}\right) = -\frac{B^2}{\mu r}, \tag{11b}$$

$$\partial_t\left(\frac{1}{2}\rho v^2 + e + \frac{B^2}{2\mu}\right) + \frac{1}{r}\partial_r\left[rv\left(\frac{1}{2}\rho v^2 + e + P + \frac{B^2}{\mu}\right)\right] = 0, \tag{11c}$$

$$\partial_t B + \frac{1}{r}\partial_r(rBv) = \frac{Bv}{r}. \tag{11d}$$

We consider the equation of state for a perfect gas, $(\gamma - 1)e = P = \rho kT/m$, where $e$ is the internal thermal energy. By injecting in equations (11a) – (11d) and boundary conditions (1) the new self-similar variables (2), we obtain the following system of Ordinary Differential Equations (ODEs) and boundary conditions,

$$\frac{dn}{d\xi} = \frac{n}{1-u}\left(\frac{2u}{\xi} + \frac{du}{d\xi}\right), \tag{12a}$$

$$\frac{dz}{d\xi} = \frac{-2\alpha b^2 + n(u - \alpha u^2 - 2\alpha z)}{\alpha\xi n} - \frac{b}{n}\frac{db}{d\xi} - \frac{z}{n}\frac{dn}{d\xi}$$
$$+ (1-u)\frac{du}{d\xi}, \tag{12b}$$

$$\frac{db}{d\xi} = \frac{b}{1-u}\left(\frac{2u}{\xi} - \frac{1}{\alpha\xi} + \frac{du}{d\xi}\right), \tag{12c}$$

$$\frac{du}{d\xi} = \frac{N}{\alpha\xi D}. \tag{12d}$$

$$n(1) = Q_1, \quad u(1) = Q_2, \quad z(1) = Q_3, \quad b(1) = Q_4, \tag{13}$$

where the numerator $N$ and denominator $D$, appearing in equation (12d) are given by,

$$N = (1 - 2\alpha)b^2 + n[(1-u)(1-\alpha u)u - 2(\alpha\gamma u + \alpha - 1)z] \tag{14}$$

$$D = b^2 - n((1-u)^2 - \gamma z), \tag{15}$$

For a given value of $\alpha$, $\gamma$, and $0 \le \sigma \le 1$, one can integrate the system of ODEs in equation (12a) – (12d) with the boundary conditions in equation (13) from the shock front at $\xi = 1$ to the position $0 \le \xi_p < 1$. When $\xi_p = 0$, the solution is integrated until the centre of the configuration and will be denoted 'continuous' (C solutions). When $0 < \xi_p < 1$, the solution meets one of the two singularities of the system $1 - u(\xi_p) = 0 \propto dR_p/dt - v(R_p, t) = 0$, which represents a tangential discontinuity. This solution denoted 'Discontinuous' (D solutions) takes the form of a shell between the shock front at $R_{sh}$ and the tangential discontinuity at $R_p$ pushed by a hot rarefied interior bubble. The second singularity is a magnetosonic line $D = 0$ which represents the point that satisfies $\mathcal{S} - v = (c^2 + v_A^2)^{1/2}$, where $v_A$ and $c$ are the local Alfven and sound speed, respectively. One can verify that this condition is never reached in the present azimuthal situation. However, we will see in section 4 that this condition can be met in the axial case.

### 3.1 Total energy conserved $\alpha = 1/2$

This section presents the special case $\alpha = 1/2$, that is, when the total energy of the BW is conserved. This situation implies that the exponent $\delta = 1$, so the ambient magnetic field $B_0 \propto 1/r$. This magnetic field can be generated by a constant current along the $z$-axis at the centre $r = 0$. The system of ODEs in equation (12a) – (12d) is integrated until the centre $\xi = 0$ (C solution) and has the integral of energy (see Appendix C),

$$\frac{d}{d\xi}\left[\xi^4\left(\frac{1}{2}nu^2 + \frac{nz}{(\gamma-1)} + \frac{1}{2}b^2\right)(1-u) - \xi^4\left(nz + \frac{1}{2}b^2\right)u\right] = 0 \tag{16}$$

This is an extension of the Sedov–Taylor case ($\sigma = 0$) and results from the fact that the energy of any sub-layer of the BW is conserved. In particular, the total energy of the BW is constant and the magnetic energy flux $B_{sh}^2/(2\mu)2\pi R_{sh}\mathcal{S}$ gained at the shock front is lost at the centre ($r = 0$). The energy integral in equation (16) is also discussed by Greifinger & Cole (1962). When the parameter $\sigma \to 1$ (highly magnetized plasma), using the ODEs in equation (12a) – (12d) and the energy integral in equation (16), the solution takes the form,

$$n(\xi) = 1 + \frac{4(1-\sigma)}{3} + O\left[(1-\sigma)^2\right], \tag{17a}$$

$$u(\xi) = O(1-\sigma), \tag{17b}$$

$$z(\xi) = \frac{16(\gamma-1)}{27}(1-\sigma)^3\frac{1}{\xi^4} + O\left[(1-\sigma)^4\right], \tag{17c}$$

$$b(\xi) = \left(1 + \frac{(1-\sigma)}{3}\right)\frac{1}{\xi^2} + O\left[(1-\sigma)^2\right]. \tag{17d}$$

It is interesting to notice that for a highly magnetized plasma, the density in the shell is almost homogeneous and static. Although the thermal pressure is low at the shock front, it increases from the shock to the centre as $\xi^2 nz \sim 1/\xi^2 = R_{sh}^2/r^2$. The same happens for the magnetic pressure, thus $(B/B_{sh})^2 \propto R_{sh}^2/r^2$. Also the solution in the shell stops to be correct close to the tangential discontinuity where the asymptotic expansion can be found in Greifinger & Cole (1962).

### 3.2 Total energy increasing $\alpha > 1/2$

Now, we will focus on the case $\alpha > 1/2$. In this case the solution is a D solution and the dimensionless thickness of the shell $h \equiv (R_{sh} - R_p)/R_{sh} = 1 - \xi_p$ depends on the parameters $\alpha$, $\sigma$, and $\gamma$. For a given parameter $\sigma$, the thickness $h$ decreases when $\alpha$ increases. Indeed, the piston at the tangential discontinuity goes faster and compresses the shell much more. The special case $\alpha = 1$ corresponds to a fast magnetosonic shock of constant velocity driven by a constant piston velocity. For a given parameter $\alpha$, the thickness of the shell increases with the parameter $\sigma$. This effect is due to two facts. First, the compression is less efficient at the shock front with the presence of a tangential magnetic field. Secondly, the tension of the magnetic lines slows down the propagation of the plasma behind the shock front. The asymptotic expansion of equations (12a) – (12d) in $\xi - \xi_p$ small without magnetic field ($\sigma = 0$) reads,

$$n(\xi) = n_0(\xi - \xi_p)^\nu + \cdots, \quad u(\xi) = 1 + u_1(\xi - \xi_p) + \cdots$$
$$z(\xi) = z_0(\xi - \xi_p)^{-\nu} + \cdots, \tag{18}$$

where

$$\nu = \frac{1-\alpha}{\alpha(\gamma+1)-1}, \quad u_1 = -2\frac{\alpha(\gamma+1)-1}{\alpha\gamma\xi_p}. \tag{19}$$

When the magnetic field is present ($\sigma > 0$), the asymptotic expansion changes to a new behaviour in $\xi - \xi_p$ small,

$$n(\xi) = n_0(\xi - \xi_p)^{\nu_1} + \cdots, \quad u(\xi) = 1 + u_1(\xi - \xi_p) + \cdots$$
$$z(\xi) = z_0(\xi - \xi_p)^{\nu_2} + \cdots, \quad b(\xi) = b_0 + b_1(\xi - \xi_p) + \cdots \tag{20}$$

where

$$\nu_1 = \frac{1}{2\alpha - 1}, \quad \nu_2 = -\frac{3 - 2\alpha - \gamma}{2\alpha - 1}, \quad u_1 = -\frac{2\alpha - 1}{\alpha\xi_p}, \tag{21}$$

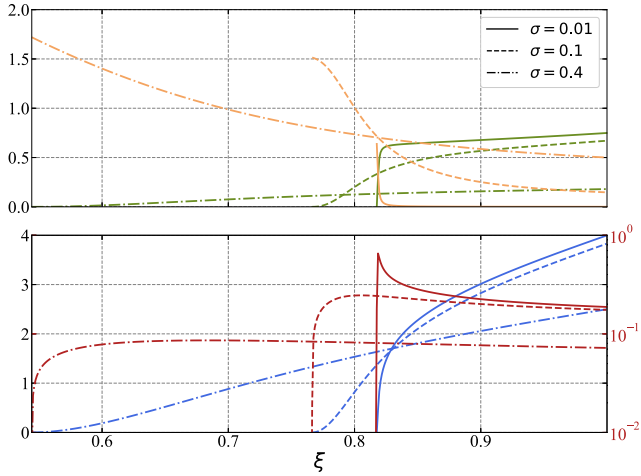

**Figure 2.** Self-similar density $n$ (blue), thermal pressure $\xi^2 nz$ (green), logarithm of temperature $\log(\xi^2 z)$ (red), and magnetic pressure $\xi^2 b^2/2$ (orange) versus the dimensionless position $\xi$ for $\alpha = 3/4$ and $\sigma = 0.01$ (solid lines), $\sigma = 0.1$ (dashed lines) and $\sigma = 0.4$ (dotted-dashed lines). The shock front is located at 1. The magnetic field orientation is azimuthal.

Fig. 2 shows the internal profiles of density $\rho/\rho_0$, thermal pressure $P/(\rho_0 \mathcal{S}^2)$, logarithm of temperature $\log(kT/(m\mathcal{S}^2))$ and magnetic pressure $B^2/(2\mu\rho_0 \mathcal{S}^2)$ for $\alpha = 3/4$ and different values of the magnetic parameter $\sigma$. As explained before, for a given decelerating parameter $\alpha$, the thickness of the shell increases with $\sigma$. For strong magnetic fields, the magnetic pressure dominates in the shell and for low magnetic fields, the thermal pressure dominates except close to the tangential discontinuity. To understand the behaviour at the inner boundary, it is important to notice that the solution bifurcates at $\sigma = 0$ (see equation 18 – 21).

When $\sigma = 0$ (no magnetic field) and for every value $1/2 < \alpha \leq 1$, the thermal pressure $\xi^2 nz$ is constant, the density $n$ goes to zero and the temperature goes to infinity ($\nu \leq 0$) at the contact discontinuity. However, as soon as the magnetic field is present ($\sigma > 0$), the behaviour changes close to the tangential discontinuity and for every value $1/2 < \alpha \leq 1$, the thermal pressure goes to zero ($\nu_1 + \nu_2 \leq 0$). Thus a transfer from thermal to magnetic energy operates in a cooling region close to the tangential discontinuity. That comes from the fact that when the field is azimuthal, there is a source term $Bv/r$ in the induction equation (see equation 11d). Thus, the magnetic field will accumulate at the lowest radius, that is, at the tangential discontinuity. This cooling process does not compress the matter as the total pressure is conserved. The cooling region thickness of high magnetic pressure decreases with $\sigma$ (see Fig. 2 top panel).

When $\sigma \to 0$, the magnetic field is only present at the piston, in a small region of order $\sigma$. Also, an additional effect is found for a certain threshold of the decelerating parameter $\alpha$. Indeed, when $\alpha > (3 - \gamma)/2$, $\nu_2 \geq 0$ and the temperature does not diverge anymore but goes to zero at the tangential discontinuity. This change of behaviour does not depend on the magnetic strength but only on the parameters $\alpha$ and $\gamma$. This can be seen in Fig. 2 bottom panel as $\alpha = 3/4 \geq (3 - \gamma)/2 = 2/3$ for $\gamma = 5/3$. The temperature goes to zero in the cooling region. Finally, when the magnetic field is low $\sigma \ll 1$, the magnetic pressure at the tangential discontinuity can be determined using equation (C5) of Appendix C. As the thermal pressure $P(R_p) = 0$, the equation reads,

$$\frac{B(R_p)^2}{2\mu} = \frac{2(2\alpha - 1)}{\alpha} \Pi(\gamma, \alpha, 0) \left(\frac{R_{sh}}{R_p}\right)^2 \rho_0 \mathcal{S}^2, \quad (22)$$

This equation illustrates the fact that the magnetic pressure at the piston ($\xi = \xi_p$) remains high due to the transfer from thermal to magnetic energy when $\sigma \to 0$. More precisely, under the latter limit, the magnetic pressure at the piston equals the thermal pressure for the same solution without magnetic field $B = 0$. However, the magnetic field is only present in an infinitely thin layer at the tangential discontinuity. Finally, when $\sigma$ increases, one needs to come back to the equation (C4),

$$P_{\text{tot}}(R_p) = \left(\frac{2(2\alpha - 1)}{\alpha} \Pi(\gamma, \alpha, \sigma) - \frac{1}{2}\sigma^2\right) \left(\frac{R_{sh}}{R_p}\right)^2 \rho_0 \mathcal{S}^2, \quad (23)$$

where $P_{\text{tot}} \equiv P + B^2/(2\mu)$. On the right side of this equation, the first term represents the dimensionless total energy increase by unit time and the second represents the input of magnetic energy at the shock front. For a given decelerating parameter $\alpha$, the difference between the two terms increases with $\sigma$. In addition, the thickness of the shell increases, and this compression of the interior also contributes to an increase in the total pressure at the piston. Thus the total pressure at the piston always increases with magnetic strength in the azimuthal case. We will see that the opposite occurs in the axial case.

## 4 AXIAL MAGNETIC FIELD

When the ambient magnetic field is along the $z$-axis, the conservation of momentum in equation (11b) and the induction in equation (11d) changes and read,

$$\rho(\partial_t v + v \partial_r v) + \partial_r \left(P + \frac{B^2}{2\mu}\right) = 0, \quad (24a)$$

$$\partial_t B + \frac{1}{r} \partial_r (r B v) = 0. \quad (24b)$$

This gives the new system of ODEs,

$$\frac{dn}{d\xi} = \frac{n}{1-u}\left(\frac{2u}{\xi} + \frac{du}{d\xi}\right), \quad (25a)$$

$$\frac{dz}{d\xi} = \frac{nu - \alpha(b^2 + n(u^2 + 2z))}{\alpha \xi n} - \frac{b}{n}\frac{db}{d\xi} - \frac{z}{n}\frac{dn}{d\xi}$$
$$+ (1-u)\frac{du}{d\xi}, \quad (25b)$$

$$\frac{db}{d\xi} = \frac{b}{1-u}\left(\frac{3u}{\xi} - \frac{1}{\alpha \xi} + \frac{du}{d\xi}\right), \quad (25c)$$

$$\frac{du}{d\xi} = \frac{N'}{\alpha \xi D}, \quad (25d)$$

$$N' = -b^2(2\alpha u + \alpha - 1)$$
$$+ n((u-1)u(\alpha u - 1) - 2(\alpha \gamma u + \alpha - 1)z). \quad (26)$$

One can notice that the two singularities of tangential discontinuity and magnetosonic line remain in the ODEs. The solution is also integrated from the shock front in equation (13) to the tangential discontinuity at $\xi_p > 0$. Yet, in the present situation, the solution can reach the condition $D = 0$ of the magnetosonic line at $\xi_p > 0$ depending on the parameters $\alpha$, $\sigma$, and $\gamma$. Thus, the solution becomes multivalued at this point ($\xi(u)$ has an extremum) and we can conclude that no physical solutions exist in this parameter space region. However, for a given value of $\alpha$ and $\gamma$, there is a special case where both the tangential discontinuity and the magnetosonic line conditions are met at the same point $\xi = 0$. As presented in Fig. 3, this line in the parameter space determines the boundary (denoted $\sigma^* \equiv \sigma^*(\alpha)$) between the physical $\sigma \leq \sigma^*$ and non-physical solutions $\sigma > \sigma^*$. The special solution at the boundary $\sigma = \sigma^*$ is a C solution.

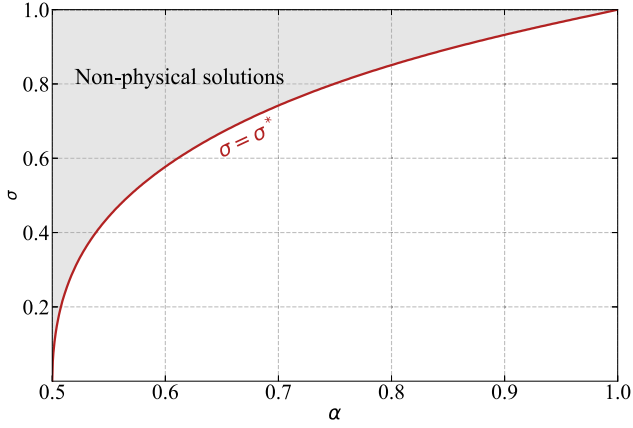

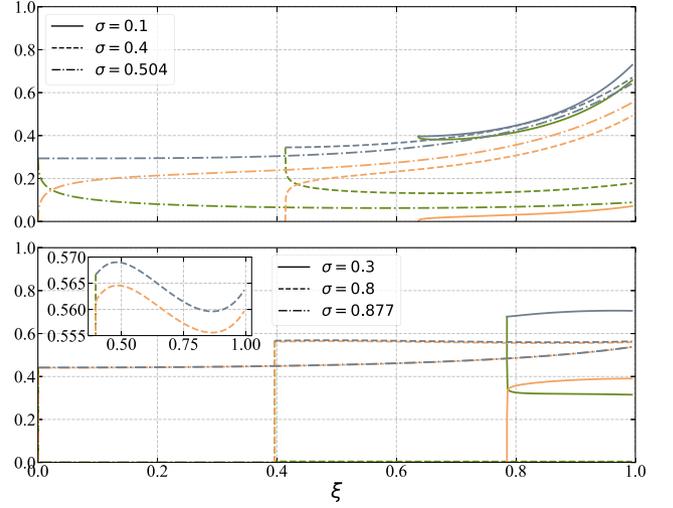

**Figure 3.** Parameter space $(\alpha, \sigma)$ in the case of axial magnetic field. The red line shows the special solution $\sigma = \sigma^*$ which delimit physical and non-physical solutions.

Indeed, when $\sigma \to \sigma^*$, the position $\xi_p$ of the tangential discontinuity goes to zero. The asymptotic expansion of equations (25a) – (25d) for $\sigma < \sigma^*$ in $\xi - \xi_p$ small reads,

$$n(\xi) = n_0(\xi - \xi_p)^{\nu_1} + \cdots, \quad u(\xi) = 1 + u_1(\xi - \xi_p) + \cdots$$
$$z(\xi) = z_0(\xi - \xi_p)^{-\nu_1} + \cdots, \quad b(\xi) = b_0(\xi - \xi_p)^{\nu_2} + \cdots$$
(27)

where

$$\nu_1 = \frac{1 - \alpha}{\alpha(\gamma + 1) - 1}, \quad \nu_2 = \frac{(1-\alpha)(2-\gamma)}{2(\alpha(\gamma+1) - 1)},$$
$$u_1 = -\frac{2(\alpha(\gamma+1) - 1)}{\alpha \gamma \xi_p}$$
(28)

On the other hand, the asymptotic expansion of the special solution $\sigma = \sigma^*$ for $\xi$ close to zero reads,

$$n(\xi) = n_0 \xi^{\nu_1} + \cdots, \quad u(\xi) = u_0 + \cdots$$
$$\xi^2 z(\xi) = z_0 \xi^{-\nu_1} + \cdots, \quad \xi b(\xi) = b_0 \xi^{\nu_2} + \cdots$$
(29)

where

$$\nu_1 = \frac{2(1-\alpha)}{\alpha(\gamma+1) - 1}, \quad \nu_2 = \frac{(1-\alpha)(2-\gamma)}{\alpha(\gamma+1) - 1}, \quad u_0 = \frac{1-\alpha}{\alpha \gamma},$$
(30)

From equation (27) to (30), we can see that unlike in the previous case of an azimuthal magnetic field, the temperature always diverges for all sets of parameters $(\alpha, \gamma, \sigma)$. Furthermore, for all values $\sigma \leq \sigma^*$, the self-similar field $\xi b$ always goes to zero at the tangential discontinuity. Fig. 4 presents the internal profiles of thermal, magnetic, and total pressure inside the shell for two different values of the decelerating parameter ($\alpha = 0.57$ for the top and $\alpha = 0.83$ for the bottom panel).

Each panel shows the profiles for three values of increasing magnetic strength parameters $\sigma$ where the highest value corresponds to $\sigma = \sigma^*$ (dotted-dashed lines). As $\sigma$ increases, the dimensionless thickness of the shell $h = 1 - \xi_p$ also increases and $\xi_p \to 0$ when $\sigma \to \sigma^*$. Also, the thermal pressure decreases, and the magnetic pressure increases in the shell but unlike in the azimuthal case, the magnetic pressure goes to zero while the thermal pressure increases to its maximal value at the piston. For low- and high-magnetic strength $\sigma$, the magnetic pressure always decreases from the shock front to the tangential discontinuity. Yet, in some range of $\sigma$, the magnetic pressure increases inside the shell with the formation of a local minimum near the shock and a local maximum near the piston (see the zoom in Fig. 4 bottom panel). For a given $\alpha$, the total pressure at the piston $P_{tot}(R_p)$ decreases when $\sigma$ increases. The reason for this is that in equation (23) the function $\Pi$ decreases for increasing magnetic strength (see Fig. 1 bottom panel). Finally, the total pressure $P_{tot}(r = 0)$ increases on the special curve $\sigma = \sigma^*$ for increasing magnetic strength.

**Figure 4.** Self-similar thermal pressure $\xi^2 nz$ (green), magnetic pressure $\xi^2 b^2/2$ (orange), and total pressure $\xi^2(nz + b^2/2)$ (grey) versus the dimensionless position $\xi$ for $\alpha = 0.57$ (top panel) and $\alpha = 0.83$ (bottom panel). The shock front is located at 1. The magnetic field orientation is axial.

## 5 NUMERICAL SIMULATIONS OF ASTROPHYSICAL MAGNETIZED SHELLS

In this section, we will present simulations of a SWB and a SNR to compare the self-similar solutions with some of the common astrophysical magnetized shells encountered in the ISM. To conserve the radial symmetry of the numerical solution as it was done for the self-similar analysis, the BW will be performed in 1D cylindrical geometry using the FLASH code (Fryxell et al. 2000). Unlike the self-similar solutions from the previous sections, the numerical solution will deviate from the similarity one as the central source $E_0 \propto t^\lambda$ and the ambient magnetic field $B_0 \propto 1/r^\delta$ power law will not meet the similarity condition $\delta \neq (2 - \lambda)/(2 + \lambda)$ (see equation 4). However, although the numerical solution will not be able to converge to a particular self-similar solution inside the parameter space, it will move in it while conserving most of the features from the self-similar solution derived in the previous sections for both azimuthal (SWB case) and axial magnetic field (SNR case). This behaviour is shown in Fig. 5, where each astrophysical object has been simulated for three different values of increasing ambient magnetic field (see later for the initial condition of ambient field). In the first case of a SWB, the BW is the result of a constant flow of stellar wind $\lambda = 1$ that propagates through an ambient azimuthal magnetic field. This field is generated by a central rotating star where $\delta = 1$, that is, the magnetic field decreases as $B_0 \propto 1/r$. Consequently, as $\lambda = 1$, the magnetic field exponent $\delta$ is greater than the value 1/3 from the self-similar solution and the magnetic field decreases more rapidly in the SWB than in the self-similar expansion. Thus, as seen in Fig. 5 top panel, the magnetic strength $\sigma$ decreases during the evolution of the simulation. At the beginning of the simulation, the decelerating parameter $\alpha$ starts from a minimum value depending on the initial conditions (this minimum is lower for increasing ambient magnetic fields of the central star). Then, the parameter $\alpha$ increases until reaching a local maximum value (this maximum is greater for increasing initial fields). Finally, $\alpha$ decreases again to reach the asymptotic solution of SWB without magnetic fields where $\alpha = 3/4$. In the second case of a SNR, the release of energy is instantaneous ($\lambda = 0$) and the ambient axial magnetic field of the ISM is homogeneous ($\delta = 0$). Unlike the previous case, $\lambda = 0$ and the exponent $\delta$ is lower than the value 1 from the self-similar solution. Thus, instead of staying on a constant value, the magnetic strength $\sigma$ increases over time as seen in Fig. 5 bottom panel. Here, the decelerating parameter $\alpha$ starts at the Sedov–Taylor solution ($\alpha = 1/2$, $\sigma = 0$). Then, $\alpha$ will always increase according to the special curve $\sigma = \sigma^*(\alpha)$ from section 4. Indeed the numerical solution always evolves as a C solution and in the axial case, $\sigma = \sigma^*$ represents the only region of C solutions.

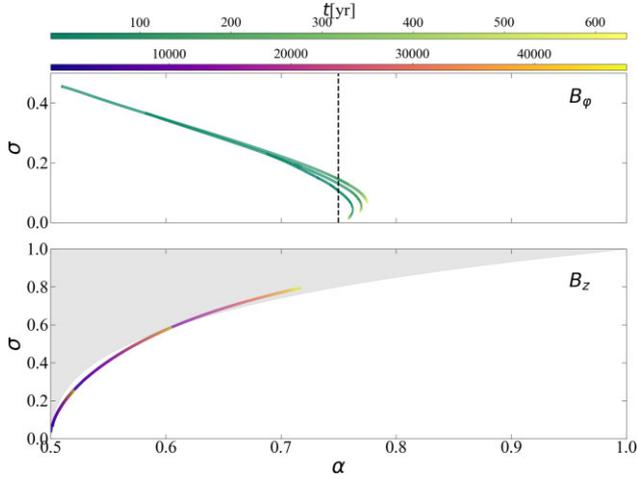

**Figure 5.** Parameter space ($\alpha$, $\sigma$) in the case of azimuthal (top panel) and axial (bottom panel) magnetic field. The time evolution of the three simulations in both cases are shown in green ($B_\varphi$ case) and purple ($B_z$ case) colour bar. The grey region in the bottom panel represents the non-physical solutions.

### 5.1 Simulation of a SWB in azimuthal ambient magnetic field

This section presents in detail the simulation of a SWB. The latter is generated from the wind of a star assumed to be in its main sequence (Chen, Zhou & Chu 2013). To simulate the propagation of a SWB in the ISM, we initialize a constant flow of stellar wind with constant velocity, density, and temperature from the centre ($r = 0$) to an arbitrary radius $R_i$ that separates the wind material from the ISM. Also, we set a boundary condition of constant input of matter at the left and an outflow condition at the right. Using these conditions, the stellar wind will bring a constant input of energy from the left that will push the BW through a static ionized ISM of constant density $\rho_0 = 10^{-23}$ g cm$^{-3}$ and temperature $T_0 = 10^4$ K (Lequeux 2004).

#### 5.1.1 Initial conditions of the stellar wind

The properties of the stellar wind are determined via the mass of the star $M_*$ and the initial radius $R_i$ where the front wind interacts with the ambient medium. First, the radius $R_*$ and the luminosity $L_*$ of the star are computed following,

$$R_* = A \left(\frac{M_*}{M_\odot}\right)^B R_\odot, \qquad L_* = C \left(\frac{M_*}{M_\odot}\right)^D L_\odot, \quad (31)$$

where $M_\odot$ and $L_\odot$ are the mass and the luminosity of the sun, respectively and $A$, $B$, $C$, and $D$ are constant depending on $M_*$

(Demircan & Kahraman 1991). Then we can deduce the mass loss rate of the star,

$$\dot{M}_* = \frac{L_*^{1.5}}{M_\odot} \left(\frac{R_*}{M_*}\right)^{2.25} G^{-1.75}, \quad (32)$$

where $G$ is the gravitational constant. Finally, we can estimate the velocity $v_W$, density $\rho_W$, and temperature $T_W$ of the wind, from the escape velocity, mass conservation, and Stephan–Boltzmann relation (Eker et al. 2015), respectively.

$$v_W = 6.2 \times 10^3 \left(\frac{M_* R_\odot}{M_\odot R_*}\right)^{1/2} \text{cm s}^{-1}, \quad (33a)$$

$$\rho_W = \frac{\dot{M}_*}{4\pi R_i^2 v_W^2}, \quad (33b)$$

$$T_W = \left(\frac{L_*}{4\pi \lambda R_*^2}\right)^{1/4}. \quad (33c)$$

For the computation of the ambient azimuthal magnetic field $B_0$ generated by the rotating star, we can determine the magnetic field $B_i = B_0(R_i)$ at position $R_i$ assuming a constant value of the magnetic field $B_*$ at the surface of the star using the equation from Usov & Melrose (1992),

$$B_i = 10^{-2} B_* \frac{R_*^2}{R_\odot R_i}. \quad (34)$$

From this point, the magnetic field decreases as $B_0 \propto 1/r$. We will perform three different simulations with different amplitudes of the surface magnetic field $B_*$. The numerical values that set the initial conditions of the SWB evolution for the three simulations are displayed in Table 1. We will neglect the magnetic field in the stellar flow ($B_0 = 0$ in the range $0 \leq r \leq R_i$). This will allow us to compare the different simulations of increases $B_*$ with the same input of energy from the stellar material.

#### 5.1.2 Results

As expected from the self-similar analysis, the magnetic field acts as a counter-pressure and slows down the propagation of the flow behind the shock front. Fig. 6 presents the density profile evolution for two different simulations of increasing initial fields $B_*$ (left panel to middle panel). The thickness of the shell increases with increasing magnetic strength. As a consequence, the tangential discontinuity slows down for increasing initial fields. Finally, as we discussed for the self-similar solution with an azimuthal magnetic field (see equation 10), the shock front velocity decreases with increasing initial fields of the star (see Fig. 6 right panel) even when the solution is not self-similar. Performing 3D simulations, García-Segura et al. (1999) studied the ambient azimuthal magnetic field effect on the SWB expansion generated from the wind of asymptotic giant branches. In agreement with our results, the latter field slows down both the forward shock front and the flow behind it. Also, as the SWB is spherical in their simulations, the flow in the direction of the magnetic field symmetry axis is not influenced and propagates faster which gives an 'eye-like' asymmetric shape to the bubble. In the model of Chevalier & Luo (1994), a thin shell is pushed by a hot interior bubble in cylindrical geometry. The azimuthal magnetic field also prevents the expansion in the direction perpendicular to the symmetry axis of the field. The model is then applied to SWB such as the nebula around SN 1987 A (Wampler et al. 1990).

Fig. 7 shows the internal structure of the BW with azimuthal magnetic field $B^* = 10$ G at $t = 2.5$ yr (top panel) and $t =$

**Table 1.** List of the simulation parameters in SWB configuration.

| Simulation | $M_*[M_\odot]$ | $\dot{M}_*[M_\odot\,\mathrm{yr}^{-1}]$ | $v_w[\mathrm{km/s}]$ | $\rho_w[\mathrm{g/cm}]$ | $T_w[K]$ | $R_i[R_\odot]$ | $B_*[G]$ | $B_i[G]$ | $\rho_0[\mathrm{g/cm}]$ | $T_0[K]$ |
|---|---|---|---|---|---|---|---|---|---|---|
| I | 40 | $1.7\,10^{-7}$ | $1.4\,10^3$ | $3.9\,10^{-23}$ | $4.6\,10^4$ | $5.7\,10^3$ | 10 | $1.4\,10^{-3}$ | $10^{-23}$ | $10^4$ |
| II | 40 | $1.7\,10^{-7}$ | $1.4\,10^3$ | $3.9\,10^{-23}$ | $4.6\,10^4$ | $5.7\,10^3$ | 30 | $3.8\,10^{-3}$ | $10^{-23}$ | $10^4$ |
| III | 40 | $1.7\,10^{-7}$ | $1.4\,10^3$ | $3.9\,10^{-23}$ | $4.6\,10^4$ | $5.7\,10^3$ | 60 | $6.5\,10^{-3}$ | $10^{-23}$ | $10^4$ |

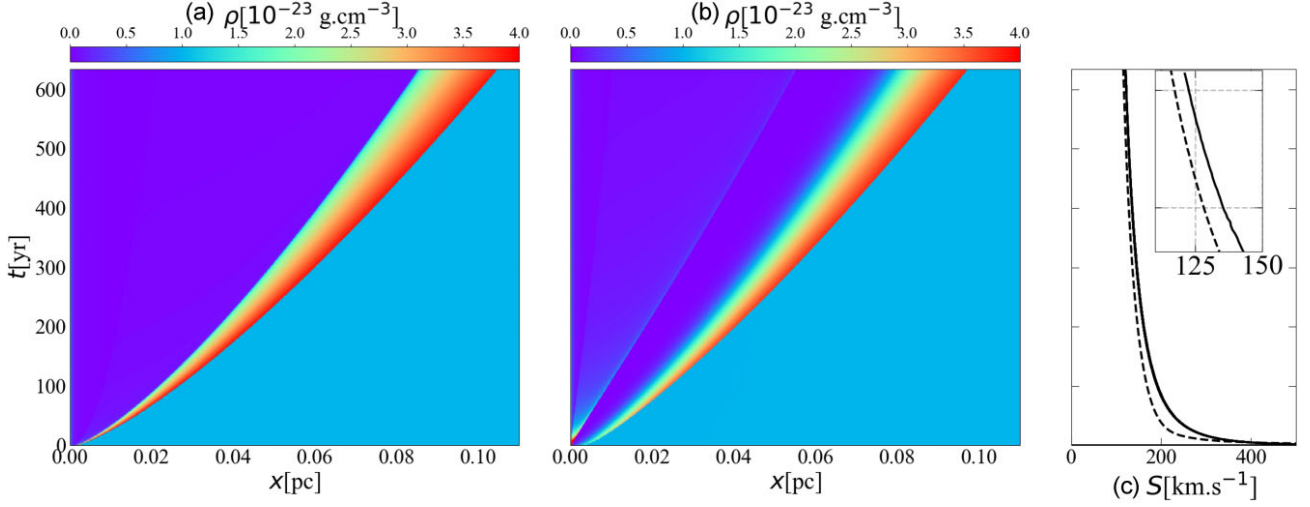

**Figure 6.** Evolution over time of plasma density profiles for the simulations $B_* = 0$ (a) and (b) $B_* = 60\,\mathrm{G}$. Panel (c) shows the time evolution of the shock front velocity for $B_* = 0$ (solid) and $B_* = 60\,\mathrm{G}$ (dashed).

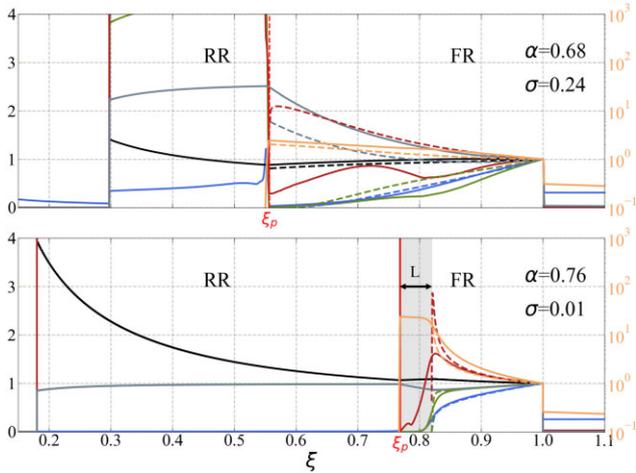

**Figure 7.** Density (blue), velocity (black), pressure (green), temperature (red), magnetic field (orange), and total pressure (grey) versus the dimensionless position $\xi$ for the simulation (solid lines) and self-similar solution (dashed lines). The forward shock front is located at 1 and the contact discontinuity at $\xi_p$. The quantities are normalized to their value at the shock front. The surface magnetic field is $B_* = 10\,\mathrm{G}$. The simulation time is $t = 2.5\,\mathrm{yr}$ (top) and $t = 631.5\,\mathrm{yr}$ (bottom).

631.5 yr (bottom panel). All the physical quantities are normalized to their values at the forward shock front ($\xi = 1$). The density $\rho_{sh} = 3.22 \times 10^{-23}\,\mathrm{g\,cm^{-3}}$ and $3.87 \times 10^{-23}\,\mathrm{g\,cm^{-3}}$, velocity $v_{sh} = 3.08 \times 10^7\,\mathrm{cm\,s^{-1}}$ and $8.95 \times 10^6\,\mathrm{cm\,s^{-1}}$, pressure $P_{sh} = 8.32 \times 10^{-9}\,\mathrm{Pa}$ and $1.08 \times 10^{-9}\,\mathrm{Pa}$, temperature $T_{sh} = 3.11 \times 10^6\,\mathrm{K}$ and $3.37 \times 10^5\,\mathrm{K}$, magnetic field $B_{sh} = 3.87 \times 10^{-4}\,\mathrm{G}$ and $6.87 \times 10^{-6}\,\mathrm{G}$ at the shock front in top and bottom panel, respectively. The numerical solution (solid curves) is compared with the self-similar solution of Section 3 (dashed curves) where the parameters $\alpha$ and $\sigma$ are given from the values computed in the simulation. The internal structure can be divided into three parts. The first one called Forward Region (FR) is bounded by the shock front and the tangential discontinuity at $\xi = \xi_p$. This is the magnetized shell resulting from the accumulation of interstellar material discussed in Section 3. The second is the Reverse Region (RR) between the tangential discontinuity and the reverse shock. This region is composed of hot shocked stellar wind where the magnetic field is null. Finally, the central region (near $r = 0$) is the source of stellar wind. This upstream region is cooler and carries mostly kinetic energy. The analytical solution in the central region can be derived from equation (11a) to (11d). By considering a stationary flow, using the conservation of entropy $P\rho^{-\gamma} = const.$ and assuming that the thermal energy is negligible compared to the kinetic energy, the solution reads,

$$\rho(r) = \rho_W \frac{R_i}{r}, \tag{35a}$$

$$v(r) = \sqrt{v_W^2 - \frac{P_W}{\rho_W}\frac{2\gamma}{\gamma-1}\left[\left(\frac{R_i}{r}\right)^{\gamma-1} - 1\right]}, \tag{35b}$$

$$T(r) = T_W \left(\frac{R_i}{r}\right)^{\gamma-1}. \tag{35c}$$

$$B(r) = 0. \tag{35d}$$

As one can see in the top panel (early stage), the effect of the ambient magnetic field is strong $\sigma = 0.24$ and increases the thickness of the FR. As the magnetic field increases from the shock front to the tangential discontinuity, the thermal energy is converted to magnetic energy. This transfer acts as a cooling inside the shell in agreement with the self-similar solution (dashed lines). The temperature drops and the magnetic field increases close to $\xi = \xi_p$ to form a fourth region denoted the Cold Magnetic Region (CMR). In this rarefied region, the cooling operates at almost constant density meaning that the thermal pressure also drops at the tangential discontinuity. However, the magnetic field is strong in this region which implies that the total pressure (purple line) is still high enough to ensure its continuity at the interface between the hot non-magnetized RR and the highly magnetized CMR. As the BW propagates in the decreasing ambient field $B_0 \propto 1/r$, the magnetic strength $\sigma$ decreases, and the maximum of the magnetic field also decreases in the CMR. Thus, the numerical solution experiences a slow transition from a magnetized solution to the self-similar solution without magnetic fields with an additional CMR region. This can be seen in the bottom panel of Fig. 7, where $\sigma = 0.01$ and the CMR region has a thickness equal to $L$. At this time, the thickness of the FR has decreased and the solution is in good agreement with the self-similar solution $\alpha = 3/4$ and $\sigma = 0.01$ (dashed lines). Also, the relative thickness of the CMR compared to the FR is decreasing with time to finally form a thin layer $L \ll (R_{sh} - R_p)$ at the tangential discontinuity. Although the magnetic field decreases in the CMR, the compression between the ambient magnetic field and the maximal magnetic field in the CMR always increases as expected from the self-similar analysis. For the self-similar solution, the maximal magnetic field at the tangential discontinuity is expected to reach the value $B_p = B(R_p) = 1.246\sqrt{\mu\rho_0}\mathcal{S}$ when $\sigma \to 0$ (see equation 22). This asymptotic regime is recover in the SWB solution and can be verified at $t = 631.5$ yr ($\sigma = 0.01 \ll 1$), where $B_p/(\sqrt{\mu\rho_0}\mathcal{S}) = 1.225$. At the early stage of the expansion ($t = 2.5$ yr), the ambient magnetic field is $B_0 = 120\,\mu$G and $\sigma = 0.24$. At this time, the corresponding maximal field in the CMR is $B_p = 940\,\mu$G, so the compression is around 8. At $t = 631.5$ yr, the ambient magnetic field is $B_0 = 10\,\mu$G and $\sigma = 0.01$. The corresponding value of $B_p = 320\,\mu$G means that the compression is about 32 and the magnetic field is still high in the CMR when $\sigma$ goes to zero.

## 5.2 Simulation of a SNR in axial ambient magnetic field

The SNR results from the release of a huge amount of energy in a small volume in the ISM. After a ballistic stage of the SNR where the velocity is constant, the ISM material swept by the shock front becomes comparable to the mass of the ejecta from the star. Thus, the SNR enters the Sedov–Talor stage where the total energy is conserved. In this section, we will see that this property is not true anymore when the SNR encounters an ambient axial magnetic field. Indeed if the shock velocity becomes comparable to the Alfven velocity, then, the parameter $\sigma$ is no longer negligible, and the total energy increases by accumulating magnetic energy from the ISM as the SNR continues to expand. It is also common that the SNR experiences magnetic effects after entering the radiative stage when the energy lost by radiation is comparable to the total energy of the shell (Petruk, Kuzyo & Beshley 2015; Petruk et al. 2018; Badjin & Glazyrin 2021). However, we will not take this effect into account in the following.

### 5.2.1 Initial condition

To simulate the propagation of the SNR in a magnetized ISM, we will generate an initial condition of a Sedov–Taylor solution in 1D cylindrical geometry aligned with the constant ambient magnetic field axis. Using a cylindrical geometry will allow us to compare the SNR with the self-similar model, and conserve the radial symmetry of the problem. This would not be the case in spherical geometry with an axial magnetic field. The Sedov–Taylor solution ($\alpha = 1/2$ and $\sigma = 0$) can be directly computed using an initial cylindrical energy of the supernova $E_{cyl}$, an ambient density $\rho_0 = 10^{-25}\,\mathrm{g\,cm^{-3}}$, which is typical for H II diffused ionized medium (Lequeux 2004) or also denoted hot low-density medium (McKee & Ostriker 1977) and an initial time $t_0 = 500$ yr which represents the end of the free-expansion (ejecta-dominated). By using an initial condition of Sedov–Taylor, we neglect the effect of the SNR ballistic stage. Thus, the radius of the SNR reads (Sedov 1959),

$$R_{sh} = \zeta \left(\frac{E_{cyl}}{\rho_0}\right)^{1/4} t^{1/2} \qquad (36)$$

where $\zeta = 1.153$ for $\gamma = 5/3$. In order to compare the initial energy of the simulation with the energy $E_{sph} = 10^{51}$ erg of the spherical SNR we use the relation $E_{cyl} = E_{sph}/L_{SNR} = 7,2 \cdot 10^{30}$ erg, where $L_{SNR} = 30$ pc is the characteristic scale of the remnant. Thus, the SNR in 1D cylindrical geometry will represent a remnant similar in size and shock velocity to the spherical SNR. Unlike the case of the self-similar analysis, the ambient axial magnetic field is assumed to be constant. This case represents a particular situation in axial fields. Indeed, the magnetic field is going to be proportional to the density in both the shell and the ISM. From the conservation of mass in equation (11a), the induction equation in the axial case in equation (24b) and the boundary condition (1), $B \propto \rho$ is a solution to the Cauchy problem when the ambient field is constant and the numerical solution will always fulfill $B/\rho = cte$ inside the shell which is not the case for the self-similar solution. We performed three simulations where the strength of the ambient magnetic field was $B_0 =$ 10, 30, and 60 $\mu$G.

### 5.2.2 Results

Fig. 8 presents the density diagram of the SNR without magnetic field (left panel) and with magnetic field $B_0 = 60\,\mu$G (middle panel). Similarly to the previous azimuthal case, the tension of axial magnetic lines decelerates the fluid motion behind the shock front. In presence of a magnetic field, the matter is more diluted inside the SNR than in the case without magnetic fields (Sedov–Taylor solution). Finally, an important result predicted by the self-similar analysis (see equation 10) is that the shock front goes faster with an axial magnetic field than without (Fig. 8 right panel). Also, in the present situation, the total energy of the BW grows as $dE/dt = 2\pi R_{sh}\mathcal{S}B_0^2/(2\mu)$, which is the ISM magnetic energy swept by the BW. By gaining energy, the BW will drive a shock that goes even faster, although the flow behind the shock goes slower due to the ISM magnetic counter-pressure. Both features of forward shock acceleration and shocked flow deceleration are also recovered in the simulations of SNR (Mineshige & Shibata 1990; Wu & Zhang 2019), SWB (van Marle et al. 2015), Galactic Centre Radio Bubbles (Zhang, Li & Morris 2021), and superbubbles (Mineshige, Shibata & Shapiro 1993) in the direction perpendicular to the ambient magnetic field axis. In the laboratory experiment of Vlases (1964a), Vlases (1964b), a cylindrical MHD shock wave is generated using an 'inverse pinch' machine. The BW is produced by the discharge of a central capacitor bank and propagates in a pre-ionized argon gas chamber imbibed in an axial magnetic field. The authors show that the shock speed increases from 10 to 15 km s$^{-1}$ when the magnetic strength increases from $\sigma = 0.2$ to $\sigma = 0.45$. In the experiment of Mabey et al. (2020), a spherical laser-induced BW is propagating in an axial ambient magnetic field. The flow behind the shock is decelerated in the direction perpendicular to the field as predicted by our model.

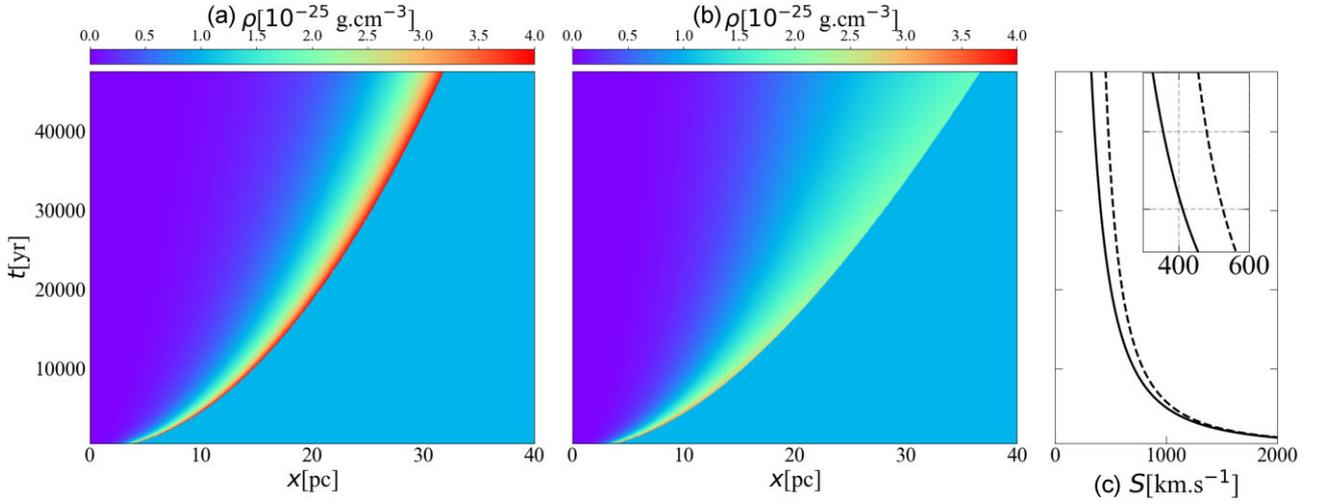

**Figure 8.** Evolution over time of plasma density profiles for the simulations $B_0 = 0$ (a) and (b) $B_0 = 60 \mu G$. Panel (c) shows the time evolution of the shock front velocity for $B_0 = 0$ (solid) and $B_0 = 60 \mu G$ (dashed).

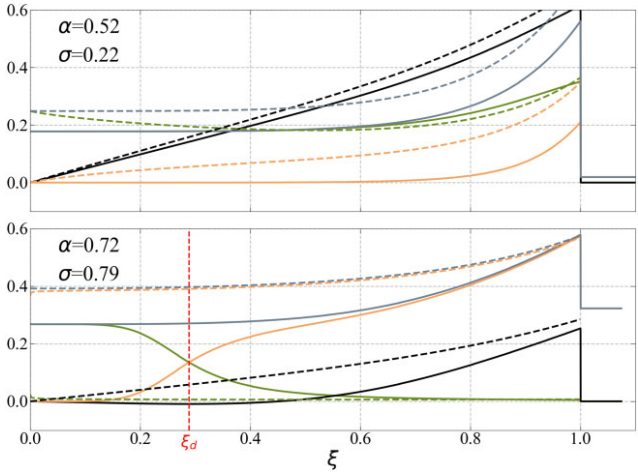

**Figure 9.** Velocity (black), thermal pressure (green), magnetic pressure (orange), and total pressure (grey) versus the dimensionless position $\xi$ for the simulation (solid lines) and self-similar solution (dashed lines). The forward shock front is located at 1. The quantities are normalized to $S$ and $\rho_0 S^2$ for the velocity and the pressures, respectively. The ambient magnetic field is $B_s = 60 \mu G$. The simulation time is $t = 1000$ yr (top) and $t = 50\,000$ yr (bottom).

Fig. 9 shows the normalized internal profiles of thermal, magnetic and total pressures as well as the fluid velocity for an ambient field $B_0 = 30 \mu G$ at $t = 1000$ yr (top panel) and at $t = 50\,000$ yr (bottom panel). The shock is propagating at $S = 2430 \,\mathrm{km\,s^{-1}}$ in the top and $S = 677 \,\mathrm{km\,s^{-1}}$ in the bottom panel. The numerical solution (solid lines) is compared to the self-similar solution (dashed lines) on the curve $\sigma = \sigma^*$. This self-similar C solution is plotted using the computed parameters $\alpha$ from the simulation and finding the corresponding $\sigma^*(\alpha)$ from the special curve in Fig. 3. Unlike the SWB, the solution is regular from the shock front to the centre of the SNR (C solution). At early times (top panel), the magnetic strength is low $\sigma = 0.22$, the decelerating parameter increases from $\alpha = 0.5$ (Sedov–Taylor solution) to $\alpha = 0.52$ and the shock front gains speed. However, unlike the self-similar C solution found on the special curve $\sigma = \sigma^*$, the pressure $P_c$ at the centre $r = 0$ does not always increase when $\sigma^*$ increases. In Fig. 10, the central pressure normalized to the shock front is plotted for different simulations of ambient fields. In all the different cases, the curve is following the same path where the normalized central pressure decreases to reach a minimum value at $\sigma_{min} = 0.5$ and increases from that point when the magnetic strength increases. Notice that the dimensionless pressure $P_c/\rho S^2$ can be recover from the relation $P/\rho S^2 = Q_1(\sigma)Q_3(\sigma)P/P_{sh}$ (see equation 1). Thus, the position $\sigma$ corresponding to the minimum value of $P_c/\rho S^2$ is different from $P_c/P_{sh}$ and corresponds exactly to the position $\sigma'_{min} = 0.6$ where the numerical solution crosses the special curve $\sigma = \sigma^*$ in the parameter space (see Fig. 5). Indeed, the present case is different from the self-similar solution because the ambient magnetic field is constant. Moreover, as mentioned before, the constant field implies that the field $B \propto \rho$. Thus the solution will not follow exactly the same curve as the self-similar C solution $\sigma = \sigma^*$ (see Fig. 5 bottom panel). To better understand the physical mechanism behind the central pressure variations, let's come back to the internal profiles (see Fig. 9). First, unlike the case of an azimuthal field, there is no additional source term $Bv/r$ in the induction equation (24b) and the magnetic pressure is only increasing near the shock front while the magnetic pressure is null at the centre of the SNR. As we have seen in the self-similar solution (dashed lines), the magnetic energy is progressively replaced by thermal energy near the centre. The same also occurs inside the SNR but as $B \propto \rho$, the magnetic pressure will fall to zero far away from the centre. This will slowly form a tangential surface at position $\xi = \xi_d$ that follows the expansion of the forward shock. Then, because the magnetic pressure continues to increase near the shock front ($\sigma$ increases with time), the ratio between the average total pressure at the right of the surface $\xi_d$ and the left is increasing. Thus, the surface decelerates until it becomes static and starts to go in the opposite direction toward the centre. At this moment, the local velocity goes to zero and becomes negative (see Fig. 9 bottom panel). Finally, the central region $\xi \leq \xi_d$ is compressed and the central pressure $P_c$ will increase from this point (see Fig. 10).

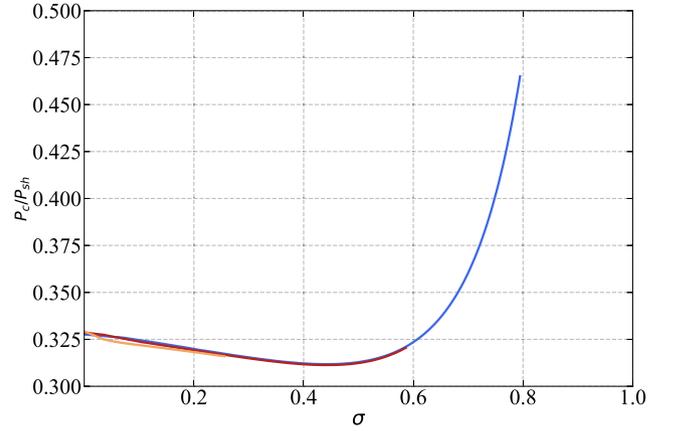

**Figure 10.** Pressure at the centre of the explosion for the simulation $B_0 = 10$ (orange), 30 (red), and 60 $\mu G$ (blue) in function of $\sigma$. The pressure is normalized at the shock front value $P_{sh}$.

## 6 CONCLUSIONS AND DISCUSSIONS

In this article, a cylindrical BW immersed in an ambient magnetic field has been studied for both cases of azimuthal and axial tangential fields. Using a self-similar form of the ambient magnetic field $B_0 \propto 1/r^\delta$, where $r$ is the distance from the centre and assuming a central energy source of the BW in the form $E \propto t^\lambda$, where $t$ is the time, a new class of self-similar solutions are found when the relation $\delta = (2 - \lambda)/(2 + \lambda)$ is fulfilled. Thus the radius $R_{sh}$ of the BW follows a power-law expansion $R_{sh} \propto t^\alpha$ where the exponent $\alpha \in [1/2, 1]$ is the decelerating parameters. We denoted three peculiar cases, $\alpha = 1/2$ where the total energy of the BW is conserved, $\alpha = 1$ which corresponds to a shock driven by a constant piston velocity, and $\alpha = 3/4$ which represents a constant input of energy from the central source. A second parameter called the magnetic strength is introduced as $\sigma \equiv M_A^{-1} \in [0, 1]$, where $M_A = \sqrt{\mu \rho_0} \mathcal{S}/B_0$, $\rho_0$, $\mu$, and $\mathcal{S}$ are the Alfven Mach number, the ambient density, the permeability of the vacuum and the forward shock velocity respectively. When $\sigma = 0$, the BW is not magnetized, and in the strong magnetic limit $\sigma = 1$, the fast magnetosonic shock is not compressible anymore. Usually, the solution denoted Discontinuous (D) solution is a shell where the outer boundary is the forward shock and the inner boundary is a tangential discontinuity that separates the dense shell from the hot BW interior region. The results show that the thickness of the shell increases and the tangential discontinuity velocity decreases with increasing magnetic strength $\sigma$, irrespective of the magnetic field orientation. Those features have already been discussed in the literature to explain the observation of asymmetric morphology in both SWB and SNR (Wampler et al. 1990; Mineshige & Shibata 1990; Mineshige et al. 1993; Chevalier & Luo 1994; García-Segura et al. 1999; van Marle et al. 2015; West et al. 2016; Wu & Zhang 2019; Zhang et al. 2021). Indeed the ambient magnetic pressure tends to decelerate the expansion in the direction perpendicular to the magnetic field. However, the present study reveals a more complex flow where the internal structure and the shock front evolution depend on the magnetic field geometry. The magnetic energy will accumulate at the tangential discontinuity or near the forward shock if the field is azimuthal or axial, respectively. This will imply different behaviour regarding the BW dynamics. Indeed, in the azimuthal case, there is a conversion from thermal to magnetic energy that acts as a cooling in the region close to the tangential discontinuity. The cooling does not compress this highly magnetized region as the total pressure (magnetic + thermal) remains high. When the decelerating parameter $\alpha \geq (3 - \gamma)/2$, where $\gamma$ is the polytropic index of the gas, the temperature even goes to zero at the tangential discontinuity. Finally, the forward shock is found to go slower (azimuthal case) or faster (axial case) than the situation without an ambient magnetic field. For a given $\alpha$ in the axial case, all the values of the magnetic strength $\sigma$ are not allowed. Indeed, there is a boundary $\sigma = \sigma^*(\alpha)$, where the solution is regular from the forward shock to the centre and is denoted continuous (C) solution. The parameter region $\sigma < \sigma^*$ are all D solutions and when $\sigma > \sigma^*$, there are no physical solutions allowed. Secondly, the self-similar solution is compared with cylindrical BW using astrophysical plasma conditions and ambient magnetic fields for the cases of SWBs and SNRs. In the case of a SBW, a central energy source of stellar wind pushes the BW which is immersed in an azimuthal ambient magnetic field $B \propto 1/r$ generated from the rotating central star. Although the solution is not self-similar anymore ($\delta = 1 \neq (2 - \lambda)/(2 + \lambda) = 1/3$), it is found to be in good qualitative agreement with the self-similar solution where the formation of a Cold Magnetized Region (CMR) is recovered close to the tangential discontinuity and the forward shock goes slower in presence of an ambient magnetic field. In the case of a SNR, the BW is initialized from the Sedov–Taylor solution (Sedov 1959) where the ambient ISM has an axial magnetic field. During the expansion of the SNR, the magnetic strength increases, and the numerical solution follows the special curve $\sigma = \sigma^*$ from the parameter space. However, there is the formation of an additional pseudo-tangential surface since the case of a constant axial magnetic field is not self-similar and $B \propto \rho$.

Although the detailed features of the BW presented in this work are difficult to investigate using observational methods, the similarity properties of the flow make it a suitable configuration to study magnetized SWB and SNR in laboratory conditions. The role of an ambient axial magnetic field in bilateral SNR morphologies has been investigated in the LULI facility by irradiating a graphite pin with a 45 J energy and 1 ns duration laser pulse (Mabey et al. 2020). The acceleration of the forward shock front when increasing the magnitude of the axial magnetic field has been confirmed in the experiments of Vlases (1964b), Vlases (1964a). Moreover, the study of SWB in the laboratory requires sustaining a constant flux of energy input at the centre of the BW. In the experiment by Burdiak et al. (2017), the Magpie pulsed-power device (Mitchell et al. 1996) creates the central kinetic plasma source by a 1.4 MA and 240 ns rise-time current-driven ablation of fine aluminum wires. The current also provides an azimuthal magnetic field $B_0 \propto 1/r$ of about 5 T which could be relevant to study SWB in azimuthal field.

Finally, when entering the radiative stage, the BW starts to lose a significant amount of energy by radiation. Thus, cooling effects will slow down the expansion of the BW and compress the matter behind the shock front (Blondin et al. 1998; Bandiera & Petruk 2004; Grun et al. 1991). Although the magnetic field is found to prevent high compression due to the tension of magnetic lines, it would be interesting to understand how the flow will react to the cooling effect in different magnetic field geometries.

## ACKNOWLEDGEMENTS

QM thanks the Czech Science Foundation under the grant number GACR 20-19854S titled 'Particle Acceleration Studies in Astrophysical Jets'. AG also thanks the project Advanced research using high-intensity laser produced photons and particles (ADONIS) (CZ.02.1.01/0.0/0.0/16_019/0000789) from European Regional Development Fund.

## DATA AVAILABILITY

The data underlying this article will be shared on reasonable request to the corresponding author.

## APPENDIX A: IDEAL MHD JUMP CONDITIONS

The ideal MHD jump conditions at the boundary between the ISM (index '0') and the shell (index 'sh') are presented for the case of a perpendicular magnetic field. Thus, the Rankine–Hugoniot equations read,

$$\rho_{sh} v_{sh} = \rho_0 \mathcal{S}, \tag{1a}$$

$$\rho_{sh} v_{sh}^2 + P_{sh} + \frac{B_{sh}^2}{2\mu} = \rho_0 \mathcal{S}^2 + P_0 + \frac{B_0^2}{2\mu}, \tag{1b}$$

$$\left( \frac{\gamma}{\gamma-1} \frac{P_{sh}}{\rho_{sh}} + \frac{1}{2} v_{sh}^2 + \frac{B_{sh}^2}{\mu \rho_{sh}} \right) \frac{\rho_{sh} v_{sh}}{\rho_0 \mathcal{S}}$$
$$= \left( \frac{\gamma}{\gamma-1} \frac{P_0}{\rho_0} + \frac{1}{2} \mathcal{S}^2 + \frac{B_0^2}{\mu \rho_0} \right), \tag{1c}$$

$$B_{sh} v_{sh} = B_0 \mathcal{S}. \tag{1d}$$

Then, we define the quantities $\rho_{sh} = Q_1 \rho_0$, $v_{sh} = Q_2 \mathcal{S}$, $P_{sh} = Q_3/Q_2 P_0$, and $B_{sh} = Q_4 \sqrt{\mu \rho_0} \mathcal{S}$. By definition of the Mach number $M$ and the Alfven Mach number $M_A$, we can write the two relations,

$$P_0 = \frac{\rho_0 \mathcal{S}^2}{\gamma M^2}, \tag{A2}$$

$$B_0 = \sqrt{\frac{\mu \rho_0 \mathcal{S}^2}{M_A^2}}. \tag{A3}$$

Finally, injecting equation (A2) and (A3) together with $Q_1$ to $Q_4$ into equation (A1a) – (A1d), one can find the dimensionless quantities for high Mach numbers ($M \to \infty$),

$$Q_1 = \frac{2(\gamma+1)}{\gamma - 1 + \gamma M_A^{-2} + M_A^{-2} \sqrt{\Delta}}, \tag{4a}$$

$$Q_2 = \frac{3 + \gamma - \gamma M_A^{-2} - M_A^{-2} \sqrt{\Delta}}{2(\gamma+1)}, \tag{4b}$$

$$Q_3 = \frac{\gamma(\gamma-1)}{4\gamma(2-\gamma)(\gamma+1)^2} \left[ 9 - \gamma(2-\gamma) - M_A^{-4}(2-\gamma) \right.$$
$$\left. [\gamma + \sqrt{\Delta}] + M_A^{-2} \left[ 18 - 7\sqrt{\Delta} + \gamma \left( 4 - 2\gamma - \sqrt{\Delta} \right) \right] \right], \tag{4c}$$

$$Q_4 = Q_1 M_A^{-1}, \tag{4d}$$

where

$$\Delta = \frac{(\gamma-1)^2}{M_A^{-4}} + \gamma^2 + \frac{8 - 2(\gamma-1)\gamma}{M_A^{-2}}. \tag{A5}$$

## APPENDIX B: DIMENSIONLESS ENERGIES DEFINITIONS

Let's consider the total magnetic energy swept by the BW (6). The associated dimensionless energy reads,

$$\Pi_{\text{tot}}(\gamma, \alpha, \sigma) = \int_{\xi_p}^{1} \frac{1}{2} \sigma^2 \xi^{\frac{2-3\alpha}{\alpha}} \, d\xi \tag{B1}$$

Then, we define the difference between the total magnetic energy flux of the shell and the ambient energy flux that was contained in the volume of the BW before the explosion (7). The associated dimensionless energy reads,

$$\Pi_{\text{tot}}(\gamma, \alpha, \sigma) = \int_{\xi_p}^{1} \sigma \xi b \, d\xi - \int_{0}^{1} \sigma^2 \xi^{\frac{-(1-\alpha)}{\alpha}} \, d\xi \tag{B2}$$

## APPENDIX C: ENERGY INTEGRAL

The integral of energy from ideal MHD conservative equations is determined inside the shell of the cylindrical BW ($R_p \leq r \leq R_{sh}$).

First, we introduce the total energy $E_{\text{ring}}$ contained in a ring between $r = R_1$ and $r = R_2$ of the shell,

$$E_{\text{ring}} = \int_{R_1(t)}^{R_2(t)} \left(\frac{1}{2}\rho v^2 + \frac{P}{(\gamma - 1)} + \frac{B^2}{2\mu}\right) 2\pi r \, dr \quad \text{(C1)}$$

Thus, using Leibniz integral rule, the time derivative $E_{\text{ring}}$ is obtained as,

$$\frac{d}{dt}E_{\text{ring}} = \left[2\pi r \frac{dr}{dt}\left(\frac{1}{2}\rho v^2 + \frac{P}{(\gamma - 1)} + \frac{B^2}{2\mu}\right)\right]_{R_1}^{R_2}$$
$$+ \int_{R_1(t)}^{R_2(t)} \partial_t \left(\frac{1}{2}\rho v^2 + \frac{P}{(\gamma - 1)} + \frac{B^2}{2\mu}\right) 2\pi r \, dr \quad \text{(C2)}$$

Then, using equation (11c) the equation reads,

$$\frac{d}{dt}E_{\text{ring}} = \left[2\pi r(\frac{dr}{dt} - v)\left(\frac{1}{2}\rho v^2 + \frac{P}{(\gamma - 1)} + \frac{B^2}{2\mu}\right)\right]_{R_1}^{R_2}$$
$$- \left[2\pi r v\left(P + \frac{B^2}{2\mu}\right)\right]_{R_1}^{R_2} \quad \text{(C3)}$$

This equation has several interesting limiting cases. In the case of the self-similar solution with $\alpha = 1/2$, it can be demonstrated in the same way as equation (3) that the total ring energy $E_{\text{ring}}$ of arbitrary thickness is conserved. Thus, injecting transformations (2) into equation (C3), the energy integral (16) for the C solutions is recovered.

For D solutions, the total energy $E$ of the shell can provide some properties of the dynamics. In this case, $R_1 = R_p$ and $R_2 = R_{sh}$. Injecting the transformations (2), the energy equation (3) and (A4a) – (A4d) for the flux at the shock front into (C3), one gets,

$$P_{\text{tot}}(R_p) = \left(\frac{2(2\alpha - 1)}{\alpha}\Pi(\gamma, \alpha, \sigma) - \frac{1}{2}\sigma^2\right)\left(\frac{R_{sh}}{R_p}\right)^2 \rho_0 S^2, \quad \text{(C4)}$$

where $P_{\text{tot}} \equiv P + B^2/(2\mu)$. Now, assuming the asymptotic case where the magnetic field $B$ is close to zero, the equation (C4) is reduced to,

$$\frac{dE}{dt} = 2\pi R_p \frac{dR_p}{dt} P_{\text{tot}}(R_p) \quad \text{(C5)}$$

Here, all the input of energy comes from the work of the total pressure force (thermal + magnetic) at the piston ($r = R_p$) and the inputs of internal and magnetic energy from the ambient medium at the shock are negligible.

This paper has been typeset from a T$_{\text{E}}$X/LAT$_{\text{E}}$X file prepared by the author.